

\magnification=\magstep1
\openup 1\jot
{\rightline {January 1993}}
\vskip .1 true in
\centerline {\bf DECOHERENCE FUNCTIONAL}
\centerline {\bf AND}
\centerline {\bf INHOMOGENEITIES IN THE EARLY UNIVERSE }
\vskip .5 true in
\centerline{  $^{1} $R.$\,$Laflamme\footnote{ $^ { { *} } $} {
Email:\ laf@tdo-serv.lanl.gov}
    \& $^ { 2} $A.$\,$Matacz\footnote{ $^ { { *} { *} } $} {
Email:\ amatacz@physics.adelaide.edu.au } }
\vskip .25 true in
\centerline{ $^ 1$Theoretical Astrophysics, T-6, MSB288,}
\centerline{ Los Alamos National Laboratory,}
\centerline{ Los Alamos, NM87545, }
\centerline{ U.S.A.}
\vskip .15 true in

\centerline{ $^ 1$ Department of  Physics,}
\centerline{University of Adelaide,}
\centerline{PO Box 498 Adelaide,}
\centerline{ Australia 5001.}
\vskip 1 true in

{\bf {Abstract.}} We investigate the quantum to classical transition of small
inhomogeneous fluctuations in the early Universe using the decoherence
functional of Gell-Mann and Hartle. We study two types of coarse graining;
one due to coarse graining the value of the scalar field and the other due
to summing over an environment. We compare the results with a previous
study using an environment and the off-diagonal rule proposed by Zurek. We show
that the two methods give different results.
\vfill\eject
\openup 1\jot

\proclaim 1) Introduction.

The  gravitational instability picture for  galaxy formation assumes that the
early Universe started with a very smooth background on which small density
fluctuations were superimposed.  It is these small fluctuations
which are ultimately responsible for the structure in the present Universe.
 They have been amplified by the
gravitational interaction since the beginning of the matter
dominated era and produced the galaxies we see.

In the sixties and seventies no theories were able to predict the existence of
these perturbations, they were just postulated to be there.   Zeldovich$^1$ and
Harrisson$^2$ suggested that in order to fit the observation the initial
 spectrum of
these perturbations must be roughly scale free. In 1980 Guth$^3$ proposed the
Inflationary scenario to solve the horizon, flatness and monopole problems of
the Big Bang.  This scenario asserts that the Univserse went
through a phase of very rapid expansion in its very early stage.  The Universe
would have expanded by a factor of at least $10^{28}$ in a mere $10^{-32}$
seconds.

It was soon  realised that this very rapid expansion would have very
interesting effect on fields especially the inhomogeneous part of the
 inflaton$^{4,5}$.  The state of small
inhomogeneities undergoes parametric amplification during the inflationary
period as soon as as a given mode crosses the Hubble radius.  This lead to a
scale free spectrum.  It was therefore suggested that these inhomogeneities
gave
rise to the needed density fluctuations in the early Universe.

It was argued that the quantum expectation value of the square of the field
$<\phi^2>$ can be interpreted as a statistical average of classical
perturbations.  The argument used by Guth $\&$ Pi$^2$ was that
$(<\phi^2><\pi^2_\phi>)^{1/2} >> \hbar$ and thus quantum mechanical affects
should be negligible.

Interestingly enough this consideration is not invariant under linear canonical
transformations.  To see this more clearly the Wigner function can be
calculated. In general the Wigner  function is  not  positive
and cannot represent a classical phase space density distribution but in the
case of a gaussian state it is positive.  So let's assume it can give us an
idea
of the classical phase space distribution. The Wigner function is defined as
$$
f_w(\phi, \pi_\phi) = {1\over 2\pi} \int d\Delta
           e^{i2\pi\pi_\phi\Delta} \rho(\phi-\Delta,\phi+\Delta)
\eqno(1.1)
$$
where $\rho$ is the state of the system. Figure 1 depicts the $1-\sigma$
contour of
the Wigner function for a mode k of a massless scalar field in the
Bunch-Davies vacuum.  Initially it is an ellipse rotating with frequency
$k/2\pi$ whose amplitude is adiabatic.  As soon as
the wavelength of the mode crosses the  Hubble radius the ellipse stop rotating
and gets elongated in the momentum direction. As seen from the  picture the
variance of $\phi$ and $\pi_{\phi}$ are such that
$(<\phi^2><\pi^2_\phi>)^{1/2}
>> \hbar$ but the surface of this ellipse remains $\hbar$.  Using a linear
canonical transformation so that $\tilde\phi$ and $\tilde\pi_\phi$ are in the
direction of the proper  axis of the ellipse would give
$(<\tilde\phi^2><\tilde\pi^2_\phi>)^{1/2} = \hbar$.  It is therefore difficult
 to
understand why the quantum mechanical average can be substituted by a
statistical one.

\vskip 4truein

\openup -2\jot
Figure 1. $1-\sigma$ contour of a Wigner
function for a mode of a scalar field which has crossed the Hubble radius.  The
area remains $\hbar$ even if $(<\phi^2><\pi^2_\phi>)^{1/2}>> \hbar$.
\openup 2\jot

A lot of effort has recently been focused on understanding the transition
between quantum and classical mechanics.
It has been proposed that a measure of the classicality of a system is obtained
by investigating the off-diagonal terms of the density  matrix$^{6,7}$.

In ref [8] such a criteria was used to investigate the
classicality of the  inhomogeneous quantum fluctuations in the inflationary
period.  It was shown that these fluctuations were not classical if they were
not interacting with an environment.  A simple model of an environment
represented by a single scalar field was constructed and it  was shown  that
the  off diagonal terms of the density matrix, in the configuration space
basis, decreased rapidly as soon as the
mode left the Hubble radius. The main problem with this approach is the
assumption that when the off-diagonal terms in configuration space vanish, the
system behaves classically.  The density matrix gives information about the
field at a given
instant in time but it does not indicate how a small cell in phase space
 evolves.
It tells us only how the sum of all these cells evolve. Classical behaviour
requires each small cell of phase space to evolve independently of the others,
that is, for there to be no quantum interference between different cells of
phase space.

In this paper we want to investigate a different approach to classicality,
the one  using the decoherence functional$^{9,10}$.  After introducing the
 method in
section 2 we investigate its consequences for perturbations in the
inflationary universe.  We study two types of coarse graining;  one due to
coarse graining of the value of the scalar field and the other by summing over
an environment as in ref [8]. We compare the coherence length of the
decoherence functional coarse grained  from an environment with that of the
density matrix and find striking differences. We discuss and conclude in
section 3.

\proclaim 2) The Decoherence Functional.

In recent years an approach to the quantum to classical transition
has considered not the state or eigenvalues of operators at a given
time but rather focused on histories defined by a series of
the value of fields at a different time$^{9,10,11,12}$. The idea is
that a necessary condition for a system to be thought of as classical is
that  the probability sum rule for different histories should be
obeyed.  In other words interference should vanish.  In such a case Griffith
called them consistent histories.

The tool to calculate the probability of an history is the
decoherence functional.  The fine grained decoherence functional for
histories defined
by the positions at all times can be
defined through the path integral
$$
D[h',h]=\delta(q_{f}'-q_f)
\exp i\left[S[q'(\eta)]-S[q(\eta)]\right ]\rho(q_i ',q_i,\eta_i)
\eqno (2.1)
$$
where $S$ is the action for the given history.
In  order for two histories $h',h$ to be consistent the decoherence
functional must be diagonal.  Except for very special cases,
fine grained histories will not decohere.  A possible way to
get consistent histories is to coarse grain them.

A coarse graining of this decoherence functional can be obtained
by looking at histories with approximate position or momenta,
or by summing over some field which is considered an environment.
It is the latter case which corresponds to the the decoherence studied
by Zeh$^7$ and Zurek$^8$.
A coarse graining can be defined as
$$
D^c[h',h]=\int_{h'}{\cal D} q'\int_{h}{\cal D}q \;\delta(q_{f}'-q_f)
        \exp i\left[S[q'(\eta)]-S[q(\eta)]\right]\rho(q_i ',q_i,\eta_i) .
\eqno (2.2)
$$
The path integral over $q(\eta)$ is over all paths that start at
$q_i$ at $\eta_i$, pass through the intervals
$\Delta^1 (\eta_1),  \Delta^2 (\eta_2)...,\Delta^n (\eta_n)$ at
$\eta_1,\eta_2...\eta_n$ and wind up at $q_f$ at time $\eta_f$. Similarly
for $q'(\eta)$ which goes through primed interval but end at the
same endpoint $q_f$.

We will evaluate (2.2)  for a scalar field evolving in the early universe
both for coarse graining of the field or of an environment.
A crucial question is how to model this environment.  Any realistic model
will be very complicated and hard to analyze.  However, the basic physics
should emerge from the simplest models.  Hence we use a model which can be
solved exactly: the system is a real massless scalar
field $\Phi_1$, (the inflaton), the environment is taken to be a second
massless real scalar field $\Phi_2$ interacting with $\Phi_1$ by their
gradients. This will permit us to compare our results with the
ones in ref.[8].  We consider the fields in the de Sitter phase of an
expanding Universe with scale factor $a (t) = \exp (Ht)$, where $H$ is
the Hubble constant. \par The action of system and environment is
$$
I= \int d^4x \sqrt{g} {1\over 2}
     \bigl ( (\partial_\mu \Phi_1)^2 + (\partial_\mu \Phi_2)^2
                              + 2c (\partial_\mu \Phi_1 \partial^\mu \Phi_2)
\bigr )
\eqno (2.3)
$$
where $g$ is the determinant of the background metric with line element given
by
$$
ds^2 = a^2 (-d\eta ^2 + dx_i^2)\, .
\eqno (2.4)
$$
$c$ is a constant measuring the strength of interaction between system and
environment.  We shall normalize the conformal time $\eta$ such that $\eta$
ranges between $- \infty$ and $0$ and $a = - (H \eta)^{-1}$ with $H^{-1}$ being
the Hubble radius.
\par
We will study two types of coarse graining.   The first one will consist in
summing over the field $\Phi_2$ which mimicks the environment. The second one
 will
consist of  coarse graining the value of the
field $\Phi_1$.
\par
Our Lagrangian is quadratic in the derivatives of the fields and can hence be
diagonalized using fields $\Phi_+$ and $\Phi_-$ for which the interaction term
disappears. The coherences in the quantum state between $\Phi_+$ and $\Phi_-
$ are only given by the initial conditions. For example,  we could choose
an initial state where
these coherences vanish. In this case, a pure state gives rise to a pure
state reduced density matrix when summing over one of the fields. Decoherence
of one field cannot occur by summing over the other one. We, however, suppose
that the inflaton and the environment do not form the diagonal basis. This
assumption is reasonable since any inflaton field
(whose reduced density matrix we want) will interact with gravitational
perturbations (part of the environment).
\par
We can expand the fields in harmonics in a box of fixed
comoving volume (physical volume $a^3$) and investigate a particular wavenumber
$k=(k_x^2+k_y^2+k_z^2)^{1/2}$.
As there is no coupling between modes with different $k$, we can  consider a
single wavelength and drop the index $k$ for convenience.  The Lagrangian
reduces to
$$
L(q,\dot{q},r,\dot{r},\eta)= {a^2(\eta)\over 2}
[\dot{q}^2+\dot{r}^2-k^2q^2-k^2r^2+2c(\dot{q}\dot{r}-k^2qr)] .
\eqno (2.5)
$$
For simplicity we will consider histories described by only two values of
$q$'s, the value at time $\eta_i$ and $\eta_f$.  When a hamiltonian exists we
ca
   n
rewrite the decoherence functional in the operator formalism as
$$
D(h_1,h_2)=Tr[U(\eta_f -\eta_i)P_{q_1 r_1}^{\sigma}\rho(\eta_i)
                 P_{q_2 r_2}^{\sigma}U^{\dag}(\eta_f -\eta_i)
                                            P_{q_f r_f}^{\sigma}] .
\eqno (2.6)
$$
In order to compare with the results using the density matrix, we will look for
histories
where $r$ is considered as an environment and at first $q$ is fine grained.
The two histories that we will consider are the  define by starting at either
$q_1$ or $q_2$ at $\eta_i$ and ending up at $q_f$. The decoherence functional
becomes
$$
D(h_1,h_2)=N\int dr_1dr_2dr_f
  K^*(q_f, r_f, \eta_f; q_2, r_2, \eta_i)K(q_f,r_f, \eta_f; q_1, r_1, \eta_i)
                          \rho(q_1, r_1; q_2, r_2, \eta_i)
\eqno (2.7)
$$
and the propagator $K$ is
$$
\eqalign{
K(q_f,r_f,\eta_f; q_1, r_1, \eta_i) & = \left[{ik^3\over2\pi H^2 x}\right]
\exp {i\over 2x} \left[{(q_f^2+r_f^2+2cq_f r_f)y_f \over H^2\eta_f^2}\right.
                   + {(q_1^2 +r_1^2 +2cq_1 r_1) y_i\over H^2\eta_i} \cr
& \hskip 1truein +
\left. {2k^3 (q_f q_1 +r_f r_1+cq_f r_1+cr_f q_1)\over H^2}\right] \cr}
\eqno (2.8)
$$
where
$$
x=-k^2\eta_f \eta_i\sin k\Delta + k \Delta\cos k\Delta- \sin k\Delta
\eqno (2.9a)
$$
$$
y_f=-k^3\eta_f\eta_i\cos k\Delta-k^2\eta_f\sin k\Delta
\eqno (2.9b)
$$
$$
y_i=-k^3\eta_f\eta_i\cos k\Delta+k^2\eta_i\sin k\Delta
\eqno (2.9c)
$$
and $\Delta=\eta_f-\eta_i$.
If we assume that the initial to be
$$
\psi(q,r,\eta_i)=a\exp-b[q^2 + r^2 + 2\alpha qr]
\eqno (2.10)
$$
where $b$ is complex and $\alpha$ is real, we find that
$$
D(h_1,h_2)= D\exp{i(1-c^2)\over 2x}
   \left[{y_i(q_1^2 - q_2^2)\over H^2\eta_i^2}
          + {2k^3q_f(q_1 - q_2) \over H^2}\right]
       exp-(q_1^2 A + q_2^2 A^* + q_1 q_2 B)
\eqno (2.11)
$$
where
$$\eqalign{
A=&[b^2(1-\alpha^2)+bb^*(1+c^2-2c\alpha)]/(b+b^*) \cr
B=&-2bb^*(c-\alpha)^2/(b+b^*) \cr
D=&aa^*\left({\pi\over b+b*}\right)^{1/2}{k^3 \over 2\pi H^2 x}.\cr}
\eqno (2.12)
$$
This decoherence functional predicts a certain coherence length $L_{df}$,
which is the maximum length (in configuration space squared) between histories
over which interference is not exponentially suppressed. However
to get
a better measure of the decoherence of the decoherence functional, $D_{df}$,
we should
divide $L_{df}$ by the probability width of the system $P_{df}$.
This is obtained by setting $q_1=q_2$ in (2.11) and finding the length in
configuration space squared where the probability not exponentially suppressed.
We find that
$$
D_{df}={P_{df} \over L_{df}}={A+A^*-B \over A+A^*+B}=1+{4bb^*(c-\alpha)^2
\over (1-\alpha^2)(b+b^*)^2}.
\eqno (2.13)
$$
When $D_{df}>>1$ we have significant histories decoherence. This measure
for histories decoherence
can be compared to the one used in ref[8], using the off--diagonal terms
of the reduced density matrix which is given by
$$
\rho_{red}(q_1,q_2,\eta_i)
=\int\psi(q_1,r,\eta_i)\psi^*(q_2,r,\eta_i)dr
=aa^*\left({\pi \over b+b^*}\right)^{1/2}exp-(q_1^2 f + q_2^2 f^* +  q_1 q_2 g)
\eqno (2.14)
$$
where
$$
g=-2\alpha^2bb^*/(b+b^*), \;\;\;\;\; f=(b^2(1-\alpha^2)+bb^*)/(b+b^*).
\eqno (2.15)
$$
In this case an analogous measure used was
$$
D_{dm}={P_{dm} \over L_{dm}}={f+f^*-g \over f+f^*+g}={(b+b)^2-\alpha^2(b-b^*)^2
 \over (1-\alpha^2)(b+b^*)^2}.
\eqno (2.16)
$$
This expression was analysed in ref[8] for the Bunch-Davies initial condition
which corresponds to
$$
\alpha=c,\;\;\;\;\;\;\;b={k^2 \over 2H^2\eta_i(k\eta_i+i)}.
\eqno (2.17)
$$
The limit of interest is long after Hubble crossing where $|k\eta_i|<<1$
which implies that $b\approx k^3/(2H^2) - ik^2/(2H^2\eta_i)$.
In this limit we see that $D_{dm}>>1$, however
$D_{df}=1$. Thus we have a situation where an arbitrarily large decoherence
of the configuration space density matrix corresponds to a maximally
coherent decoherence functional. The Bunch-Davies vacuum is the ground
state. If we perturb the initial coupling away from $c$ then we will have
some histories decoherence as well as density matrix decoherence. In this case
the propagator will generate an imaginary contribution to $\alpha$. This
imaginary part will modify (2.13) and (2.16) in a way that cancels the
divergence which would otherwise occur for $\alpha\neq c$ ($\alpha$ real)
in the limit
$k\eta_i \rightarrow 0$. In this case there may be closer relationship
between the two decoherence measures.
We can
get an idea of the relative strengths of the two decoherence measures
(for real $\alpha$)
by taking their ratios. We find
$$
{L_{dm}\over L_{df}}= {A+A^*-B \over f+f^*-g}
=1+ {c(c-2\alpha)bb^* \over(re\;b)^2+\alpha^2(im\;b)^2}.
\eqno (2.18)
$$

We can see that if $c=0$ the two coherence lengths agree.
This can be easily seen form
(2.7).  The integral over $r_f$ will be proportional to $\delta(r_1-r_2)$
and thus the decoherence functional is proportional to the initial density
matrix.  It is rather surprising however that turning on the interaction
from $c=0$ to $c=2\alpha$ will increase the coherence of the decoherence
functional relative to the density matrix. However for $c>2\alpha$
the coherence length of the density matrix
is larger than the coherence length of the
decoherence functional.   This shows that in this case there is no
obvious correlation between the two decoherence measures and that the
decoherence of the density matrix does not imply the decoherence of the
decoherence functional or vice-versa.

Another possibility to obtain decoherence is to coarse grain our system
in configuration space, to which we turn now. In this case the histories
 are not defined by precise
values of $q$ but by a range determined by $\sigma$ (a variance) around a given
value.  The decoherence functional can then be obtained by integrating the fine
grained one.
The $P$s are projectors on a range $2\sigma$ of the fields.  They are
rather
tedious to work with analytically.  It will be useful to keep the analytical
result simple so we use the gaussian pseudo-projectors
$$
P_{q_i r_i}^{\sigma}={1\over\pi^{1/2}\sigma}\int_{-\infty}^{\infty}dz_idr_i
  \exp\left({{-(z_i-q_i)^2\over\sigma ^2}}\right)  |z_i ,r_i\rangle\langle r_i,
 z_i |
\eqno (2.19)
$$
They are not exactly projectors as
$$
P^2\neq P
\eqno (2.20)
$$
but are a sufficiently good approximation for our purpose.
Substituting (2.19) into (2.6) we get
$$
D^c(h_1,h_2) = \int dz_1dz_2dz_3 \exp\Big[  -{(z_1-q_1)^2\over \sigma^2}
                                      -{(z_2-q_2)^2\over \sigma^2}
                                      -{(z_3-q_f)^2\over \sigma^2} \Big]
 D(h_1,h_2)
\eqno (2.21)
$$
where $D(h_1,h_2)$ is given by (2.11). We choose the Bunch-Davies initial
condition (2.17) for (2.11). This is the most natural initial state to choose,
it considerably simplifies the algebra and it ensures by virtue of (2.13) that
any decoherence obtained will not be due to the environmental coarse grain.
We can rewrite the result in terms of $Q= q_1 +q_2$ and $\delta=q_1 -q_2$ and
 get
$$
D^c(h_1,h_2)=N\exp[a_1Q^2 + a_2\delta^2 + a_3 q_f^2 + a_4 Q\delta
                        + a_5 q_f \delta +a_6 Qq_f]
\eqno (2.22)
$$
where
$$\eqalign{
a_1=&{-1\over 2\sigma^2}
    + {1\over 4\sigma^4}{ M + M^* + 2V \over MM^* -V^2}       \cr
a_2=&{-1\over 2\sigma^2}
    + {1\over 4\sigma^4}{ M + M^* - 2V \over MM^* -V^2}       \cr
a_3=& - { V(M + M^* - 2V) \over \sigma^2 (MM^* -V^2)}         \cr
a_4=&  { M - M^* \over 2 \sigma^4 (MM^* -V^2)}                \cr
a_5=& { iV^{1/2}(M + M^* - 2V) \over \sigma^3 (MM^* -V^2)}    \cr
a_6=& { iV^{1/2}(M - M^*) \over \sigma^3 (MM^* -V^2)}         \cr}
\eqno (2.23)
$$
with
$$
\eqalign{
M=& {1\over\sigma^2} + (1-c^2)b^* + {i(1-c^2)y_i\over 2xH^2\eta^2_i}
   +{ (1-c^2)^2k^6\sigma^2 \over 4x^2H^4}                      \cr
V=& {(1-c^2)^2k^6\sigma^2 \over 4x}. \cr}
\eqno (2.24)
$$

Investigating (2.22-24) shows that coarse grained histories that are
determined by their
approximate positions at various times are not exactly consistent. Exact
decoherence is rather difficult to obtain so we investigate approximate
 decoherence.  Histories are approximatively
 consistent if
$$
|Re D(h_1,h_2)|< \varepsilon Min[Re D(h_1,h_1),Re D(h_2,h_2)]
\eqno (2.25)
$$
This only means that the off-diagonal are much smaller than its corresponding
diagonal part and thus the classical sum rules applies approximatively.
$\varepsilon$ controls how good the approximation is.
If we consider symmetrical histories ($q_1=-q_2, q_f=0$) then it is easy to
see from (2.22) that (2.25) translates mathematically as
$$
{a_1 \over a_2} << 1.
\eqno (2.26)
$$
We also want to be able to interpret
the quantum mechanical average of operators as a statistical one. This implies
that the coarse grain should be smaller than the fluctuations $(\Delta q)^2$
of the field. For the Bunch Davies vacuum (2.17), in the long after Hubble
crossing limit, $(\Delta q)^2\rightarrow H^2/k^3$ hence we require
$$
\sigma^2 << { H^2 \over k^3}.
\eqno (2.27)
$$

In general (2.23) will be very long expressions. However they simplify
greatly in the late time limit which implies that from (2.9a,2.9c)
$y_i\rightarrow -k^3\eta_i^2$ and $ x\rightarrow{-k^3\Delta \over 3}(3\eta_i
\eta_f+\Delta^2)$. We further consider the limit $\Delta\rightarrow 0,\eta_i
\rightarrow 0$ and $\eta_f\rightarrow 0$. We can take this limit while
keeping an arbitrary constant proper time interval, $\delta t$ since
 ${d\eta \over dt}=-H\eta$.  In this limit we find that
$$\eqalign{
a_1 &\rightarrow {-1 \over 2\sigma^2} + {1 \over \sigma^4}\left[{3 \over
\sigma^2} + {(1-c^2)k^3 \over H^2}\right]^{-1} \cr
a_2 &\rightarrow {-1 \over 2\sigma^2} \cr
a_3 &\rightarrow{-(1-c^2) \over \sigma^2}\left[{2H^2+(1-c^2)k^3\sigma^2 \over
 3H^2+(1-c^2)k^3\sigma^2}\right] \cr
a_4 &\rightarrow a_5 \rightarrow 0  \cr
a_6 &\rightarrow {2(1-c^2) \over \sigma^4}\left[{3 \over
\sigma^2} + {(1-c^2)k^3 \over H^2}\right]^{-1}. \cr}
\eqno (2.28)
$$

For our model in the late time limit (2.26) becomes
$$
{a_1\over a_2} \approx  1/3.
\eqno (2.29)
$$
Equation (2.29) tells us that there is weak decoherence
but not a significant amount. To get a better feel for this number it is
worth comparing it to the long before Hubble crossing limit which gives
$a_1/a_2 =1$. Thus the after Hubble crossing limit does lead to some
decoherence
but not a significant amount.

Assuming decoherent histories ($\delta=0, q=Q/2$) we find that (2.22)
becomes, using (2.28) and (2.27)
$$
D^c(h,h)\approx N\exp\left[{-2 \over 3\sigma^2} (q-q_f)^2\right].
\eqno (2.30)
$$
Thus at late times the histories would be
peaked about  $q_f\approx q$ which is exactly the behavior of the
classical motion.

\proclaim 3) Discussion and conclusion.

We see from (2.16) that the decoherence in the density matrix is due
purely to the phase of the wave-function.
This dependence on the phase is interesting since the phase can always be
changed by a point transformation on the Lagrangian. We can see this as
follows. A point transformation will transform the Lagrangian as
$$
L(\vec{q}(t),\dot{\vec{q}}(t))\rightarrow L(\vec{q}(t),\dot{\vec{q}}(t))-
{d \over dt}f(\vec{q}(t),t)
\eqno (3.1)
$$
which in turn means that the action transforms as
$$
S[q(t)]\rightarrow S[q(t)] -f(\vec{q}_f,t_f)+f(\vec{q}_i,t_i)
\eqno (3.2)
$$
This point transformation doesn't affect the classical equation of motion
because they are derived from the stationary action condition
$\delta S=S[q(t)]-S[q(t)+\delta q(t)]=0$ where $\delta q(t)$ vanishes at the
endpoints.
However from the general expression $U(\vec{q}_f,t_f;\vec{q}_i,t_i)=
N\sum_{paths}e^{iS}$ for the quantum propagator we can see that under
the transformation (3.2) the quantum propagator transforms as
$$
U(\vec{q}_f,t_f;\vec{q}_i,t_i)\rightarrow e^{-if(\vec{q}_f,t_f)}
U(\vec{q}_f,t_f;\vec{q}_i,t_i) e^{if(\vec{q}_i,t_i)}
\eqno (3.3)
$$
which in turn means that the wave function transforms as
$$
\psi(\vec{q},t)\rightarrow e^{-if(\vec{q},t)}\psi(\vec{q},t)
\eqno (3.4)
$$
Physics is generally considered invariant under the point transformation
(3.1) because expectation values of functions of $q$ and the physical
momenta $\dot{q}$ are invariant (it is important to remember that
the canonical momenta does change with (3.2)).
However the reduced density matrix of a subsystem in is not
invariant to these point transformations. A point transformation
is exactly what is being done when surface terms are dropped in a
lagrangian. Using (3.3) and (3.2) we can see that the decoherence functional
(2.7) is invariant under point transformations. This is an important difference
between the two formalisms.

In models with more general couplings we should expect decoherence of the
reduced density matrix to depend not only on the phase but also on the real
part of the exponent of the wave function.  In this case there might be a
simpler relation between the decoherence functional and the evolution of the
density matrix.

It is also interesting to investigate the influence functional for this
model.  Naively we might relate a diagonal decoherence functional with
the existence of a noise kernel in the influence functional.
Consider (2.2) where
$q\rightarrow(q,r), q'\rightarrow(q',r')$ and the $r,r'$ coordinates are
 completely coarse-grained out. In this case
 (2.2) becomes
$$
D[h',h]=\int_{h'}{\cal D} q\int_{h}{\cal D}q'\int dq_i dq'_i dq_f dq'_f\;
\delta(q_{f}-q'_f)\exp i\left[S_f [q(\eta)]-S_f
[q'(\eta)]\right]F[q(\eta),q'(\eta)]
\eqno (3.5)
$$
where $F[q(\eta),q'(\eta)]$ the influence functional is
$$
\eqalign{
 &F[q(\eta),q'(\eta)]=\int dr_i dr'_i dr_f dr'_f\;\delta(r_f-r'_f)\rho(q_i,r_i;
q'_i,r'_i,\eta_i)  \cr
&\times \int_{(r_i,r'_i,\eta_i)}^{(r_f,r'_f,\eta_f)}
{\cal D}r{\cal D}r'\exp i\left[S_f[r(\eta)]+S_i[q(\eta),r(\eta)] -S
_f[r'(\eta)]-S_i[q'(\eta),r'(\eta)]\right]. \cr }
\eqno (3.6)
$$
For our model (2.5) with the Bunch-Davies initial condition (2.17) we
find that the influence functional is
$$ \eqalign{
F[q(\eta),q'(\eta)]
& =\exp\left[{ic^2\over 2}
\int_{\eta_i}^{\eta_f}a^2 (\dot{q}'^2-k^2 q'^2)-{ic^2\over 2}\int_{\eta_i}
^{\eta_f}a^2 (\dot{q}^2-k^2 q^2)\right] \cr
& \times \exp\left[-(1-c^2)b_i q^2_i-(1-c^2)b^*_i q'^2_i -{c^2b_fb_f^*(q_f-q'_f
)^2\over b_f+b^*_f}\right]. \cr}
\eqno (3.7)
$$
A striking feature of (3.7) is the absence of a noise kernel that is typically
associated with decoherence. This is due to the very special form of
interaction we have chosen. The result is that in (3.5), there is no
exponential
suppression of widely separated histories and hence no histories decoherence.
The influence functional  will still not  have a noise
kernel even
if the initial state does not have $c=\alpha$.  This shows that the absence
of noise kernel does not imply a coherent evolution.

We have shown that the decoherence functional shows some decoherence
for the interaction given in eq.(2.3) for a wide selection of initial states.
We have also shown that there is a surprising result for the case
 $c=\alpha$
as we have already mentioned.  This case surely needs further study in order
to understand why a mixed state can lead to a maximally coherent (factorizable)
decoherence functional. We also considered the
 possibility of decoherence
after Hubble crossing though coarse graining the system field. We found
this led to weak decoherence after Hubble crossing but probably not
enough for an effective quantum to classical transition.
\vfill\eject
\proclaim  References.

\item{1.} Ya.B. Zeldovich, {\it Mon.Not.R.Astron.Soc.} {\bf 160}, 1 (1972).

\item{2.} E.R. Harrisson, {\it Phys.Rev.D.} {\bf 1}, 2726 (1970).

\item{3.} A. Starobinsky, {\it Phys.Lett.} {\bf 91B}, 99 (1980).  A. Guth, {\it
 Phys.Rev.} {\bf D23}, 347 (1981); K.Sato, {\it Mon.Not.R.Astron.Soc.} {\bf
195}
   ,
467 (1981); A.D. Linde, {\it Phys.Rev.Lett.} {\bf 108B}, 389 (1982); A.
Albrecht
\& P.J. Steinhardt, {\it Phys.Rev.Let.} {\bf 48}, 1220 (1980).

\item{4.} S.W.Hawking, {\it Phys.Lett} {\bf 115B},295 (1982).

\item{5.} A.H.Guth \& S.Y.Pi,
 {\it Phys.Rev.Lett.} {\bf 49},  1110 (1982).

\item{6.} W.Zurek, {\it Phys.Rev.} {\bf D24}, 1516 (1981).

\item{7.} E.Joos \& H. Zeh, {\it Z.Phys.} {\bf B59}, 223 (1985).

\item{8.}
R.Brandenberger, R.Laflamme \& M.Miji\'c, {\it Mod.Phys.Lett.} {\bf A5}, 2311
 (1990).

\item{9.} J.B.Hartle, in Proceedings of the 1992 Les Houches Summer School on
Gravitation and Quantizations, ed. by B. Julia, North Holland, Amsterdam
and references therein.

\item{10.} M.Gell-Mann \& J.B.Hartle, {\it Classical Equations for Quantum
Systems}, (to appear in Phys. Rev. D.).

\item{11.} R.Griffiths, {\it J.Stat.Phys.} {\bf 36}, 219 (1984).

\item{12.} R.Omnes, {\it Rev.Mod.Phys.} {\bf 64}, 339 (1992).

\bye